\documentstyle[twoside]{article}

\catcode`\@=11
\long\def\@makefntext#1{
\protect\noindent \hbox to 3.2pt {\hskip-.9pt
$^{{\eightrm\@thefnmark}}$\hfil}#1\hfill}               

\def\thefootnote{\fnsymbol{footnote}}
\def\@makefnmark{\hbox to 0pt{$^{\@thefnmark}$\hss}}    

\def\ps@myheadings{\let\@mkboth\@gobbletwo
\def\@oddhead{\hbox{}
\rightmark\hfil\eightrm\thepage}
\def\@oddfoot{}\def\@evenhead{\eightrm\thepage\hfil
\leftmark\hbox{}}\def\@evenfoot{}
\def\sectionmark##1{}\def\subsectionmark##1{}}

\oddsidemargin=\evensidemargin
\renewcommand{\thefootnote}{\fnsymbol{footnote}}

\newcounter{sectionc}\newcounter{subsectionc}\newcounter{subsubsectionc}
\renewcommand{\section}[1] {\vspace{12pt}\addtocounter{sectionc}{1}
\setcounter{subsectionc}{0}\setcounter{subsubsectionc}{0}\noindent
        {\tenbf\thesectionc. #1}\par\vspace{5pt}}
\renewcommand{\subsection}[1] {\vspace{12pt}\addtocounter{subsectionc}{1}
        \setcounter{subsubsectionc}{0}\noindent
        {\bf\thesectionc.\thesubsectionc. {\kern1pt \bfit #1}}\par\vspace{5pt}}
\renewcommand{\subsubsection}[1] {\vspace{12pt}\addtocounter{subsubsectionc}{1}
        \noindent{\tenrm\thesectionc.\thesubsectionc.\thesubsubsectionc.
        {\kern1pt \tenit #1}}\par\vspace{5pt}}
\newcommand{\nonumsection}[1] {\vspace{12pt}\noindent{\tenbf #1}
        \par\vspace{5pt}}

\newcounter{appendixc}
\newcounter{subappendixc}[appendixc]
\newcounter{subsubappendixc}[subappendixc]
\renewcommand{\thesubappendixc}{\Alph{appendixc}.\arabic{subappendixc}}
\renewcommand{\thesubsubappendixc}
        {\Alph{appendixc}.\arabic{subappendixc}.\arabic{subsubappendixc}}

\renewcommand{\appendix}[1] {\vspace{12pt}
        \refstepcounter{appendixc}
        \setcounter{figure}{0}
        \setcounter{table}{0}
        \setcounter{lemma}{0}
        \setcounter{theorem}{0}
        \setcounter{corollary}{0}
        \setcounter{definition}{0}
        \setcounter{equation}{0}
        \renewcommand{\thefigure}{\Alph{appendixc}.\arabic{figure}}
        \renewcommand{\thetable}{\Alph{appendixc}.\arabic{table}}
        \renewcommand{\theappendixc}{\Alph{appendixc}}
        \renewcommand{\thelemma}{\Alph{appendixc}.\arabic{lemma}}
        \renewcommand{\thetheorem}{\Alph{appendixc}.\arabic{theorem}}
        \renewcommand{\thedefinition}{\Alph{appendixc}.\arabic{definition}}
        \renewcommand{\thecorollary}{\Alph{appendixc}.\arabic{corollary}}
        \renewcommand{\theequation}{\Alph{appendixc}.\arabic{equation}}
        \noindent{\tenbf Appendix \theappendixc #1}\par\vspace{5pt}}
\newcommand{\subappendix}[1] {\vspace{12pt}
        \refstepcounter{subappendixc}
        \noindent{\bf Appendix \thesubappendixc. {\kern1pt \bfit #1}}
        \par\vspace{5pt}}
\newcommand{\subsubappendix}[1] {\vspace{12pt}
        \refstepcounter{subsubappendixc}
        \noindent{\rm Appendix \thesubsubappendixc. {\kern1pt \tenit #1}}
        \par\vspace{5pt}}

\newcommand{\textlineskip}{\baselineskip=13pt}
\newcommand{\smalllineskip}{\baselineskip=10pt}

\def\eightcirc{
\begin{picture}(0,0)
\put(4.4,1.8){\circle{6.5}}
\end{picture}}
\def\eightcopyright{\eightcirc\kern2.7pt\hbox{\eightrm c}}

\newcommand{\copyrightheading}[1]
        {\vspace*{-2.5cm}\smalllineskip{\flushleft
        {\footnotesize International Journal of Modern Physics B, #1}\\
        {\footnotesize $\eightcopyright$\, World Scientific Publishing
         Company}\\
         }}

\def\abstracts#1#2#3{{
        \centering{\begin{minipage}{4.5in}\baselineskip=10pt\footnotesize
        \parindent=0pt #1\par
        \parindent=15pt #2\par
        \parindent=15pt #3
        \end{minipage}}\par}}

\def\keywords#1{{
        \centering{\begin{minipage}{4.5in}\baselineskip=10pt\footnotesize
        {\footnotesize\it Keywords}\/: #1
        \end{minipage}}\par}}

\renewenvironment{thebibliography}[1]                   
        {\frenchspacing
         \ninerm\baselineskip=11pt
         \begin{list}{\arabic{enumi}.}
        {\usecounter{enumi}\setlength{\parsep}{0pt}
         \setlength{\leftmargin 12.7pt}{\rightmargin 0pt} 
         \setlength{\itemsep}{0pt} \settowidth
        {\labelwidth}{#1.}\sloppy}}{\end{list}}

\newcounter{itemlistc}
\newcounter{romanlistc}
\newcounter{alphlistc}
\newcounter{arabiclistc}

\newcommand{\fcaption}[1]{
        \refstepcounter{figure}
        \setbox\@tempboxa = \hbox{\footnotesize Fig.~\thefigure. #1}
        \ifdim \wd\@tempboxa > 5in
           {\begin{center}
        \parbox{5in}{\footnotesize\smalllineskip Fig.~\thefigure. #1}
            \end{center}}
        \else
             {\begin{center}
             {\footnotesize Fig.~\thefigure. #1}
              \end{center}}
        \fi}

\newcommand{\tcaption}[1]{
        \refstepcounter{table}
        \setbox\@tempboxa = \hbox{\footnotesize Table~\thetable. #1}
        \ifdim \wd\@tempboxa > 5in
           {\begin{center}
        \parbox{5in}{\footnotesize\smalllineskip Table~\thetable. #1}
            \end{center}}
        \else
             {\begin{center}
             {\footnotesize Table~\thetable. #1}
              \end{center}}
        \fi}

\def\@citex[#1]#2{\if@filesw\immediate\write\@auxout
        {\string\citation{#2}}\fi
\def\@citea{}\@cite{\@for\@citeb:=#2\do
        {\@citea\def\@citea{,}\@ifundefined
        {b@\@citeb}{{\bf ?}\@warning
        {Citation `\@citeb' on page \thepage \space undefined}}
        {\csname b@\@citeb\endcsname}}}{#1}}

\newif\if@cghi
\def\cite{\@cghitrue\@ifnextchar [{\@tempswatrue
        \@citex}{\@tempswafalse\@citex[]}}
\def\citelow{\@cghifalse\@ifnextchar [{\@tempswatrue
        \@citex}{\@tempswafalse\@citex[]}}
\def\@cite#1#2{{$\null^{#1}$\if@tempswa\typeout
        {IJCGA warning: optional citation argument
        ignored: `#2'} \fi}}
\newcommand{\citeup}{\cite}

\def\pmb#1{\setbox0=\hbox{#1}
        \kern-.025em\copy0\kern-\wd0
        \kern.05em\copy0\kern-\wd0
        \kern-.025em\raise.0433em\box0}

\def\fnt#1#2{\footnotetext{\kern-.3em
        {$^{\mbox{\scriptsize #1}}$}{#2}}}

\def\fpage#1{\begingroup
\voffset=.3in
\thispagestyle{empty}\begin{table}[b]\centerline{\footnotesize #1}
        \end{table}\endgroup}

\def\runninghead#1#2{\pagestyle{myheadings}
\markboth{{\protect\footnotesize\it{\quad #1}}\hfill}
{\hfill{\protect\footnotesize\it{#2\quad}}}}
\font\tenrm=cmr10
\font\tenit=cmti10
\font\tenbf=cmbx10
\font\bfit=cmbxti10 at 10pt
\font\ninerm=cmr9
\font\nineit=cmti9
\font\ninebf=cmbx9
\font\eightrm=cmr8

\def\qed{\hbox{${\vcenter{\vbox{                        
   \hrule height 0.4pt\hbox{\vrule width 0.4pt height 6pt
   \kern5pt\vrule width 0.4pt}\hrule height 0.4pt}}}$}}

\renewcommand{\thefootnote}{\fnsymbol{footnote}}        

\def\bsc{{\sc a\kern-6.4pt\sc a\kern-6.4pt\sc a}}       
\def\bflatex{\bf L\kern-.30em\raise.3ex\hbox{\bsc}\kern-.14em
T\kern-.1667em\lower.7ex\hbox{E}\kern-.125em X}

\begin{document}

\runninghead{$\quad$~A.~M.~R.~Cadilhe,~M.~L.~Glasser~and~V.~Privman}{Exact~solutions~of~low-dimensional~reaction-diffusion~systems~$\quad$}

\normalsize\textlineskip
\thispagestyle{empty}
\setcounter{page}{1}

\copyrightheading{}                     

\vspace*{0.88truein}

\fpage{1}
\centerline{\bf EXACT SOLUTIONS OF LOW-DIMENSIONAL}
\vspace*{0.035truein}
\centerline{\bf REACTION-DIFFUSION SYSTEMS}
\vspace*{0.37truein}
\centerline{\normalsize Ant\'onio M. R. Cadilhe, M. Lawrence
Glasser and Vladimir Privman}
\vspace*{0.072truein}
\centerline{\footnotesize\it Department of Physics, Clarkson University}
\baselineskip=10pt
\centerline{\footnotesize\it Potsdam, New York 13699-5820, USA}

\vspace*{0.21truein}
\abstracts{We briefly review some common diffusion-limited reactions
with emphasis on results for two-species reactions with 
anisotropic hopping.
Our review also covers single-species reactions.
The scope is that of providing reference and general discussion
rather than details of methods and results. Recent exact results for a
two-species model with anisotropic hopping and with `sticky'
interaction of like particles, obtained by a novel method which
allows exact solution of certain single-species and two-species
reactions, are discussed.}{}{}

\vspace*{10pt}
\keywords{diffusion-limited reactions, exact solutions, hopping
anisotropy}


\vspace*{1pt}\textlineskip      
\section{Introduction}    
\vspace*{-0.5pt}
\noindent
There has been much interest in the study of nonequilibrium
systems through representation by simple stochastic
models.\citeup{41}$^{\hbox{\footnotesize -}}$\citeup{85} In
particular, low-dimensional 
diffusion-limited reactions are tractable and in many cases,
exactly solvable, examples of nonequilibrium
systems. Since the early works of Ovchinnikov and Zeldovich,\citeup{1}
Toussaint and Wilczek,\citeup{79} and
Torney and McConnell\citeup{78} who
pointed out the importance of fluctuations of particle
density in the kinetics of low-dimensional reactions, a host of results
have been published.\citeup{41}$^{\hbox{\footnotesize -}}$\citeup{71}
Despite their simplicity, common
diffusion-limited reactions can describe complex phenomena.
However, only recently attention has been turned to the effect of
biased diffusion on these
reactions.\citeup{21}$^{\hbox{\footnotesize
-}}$\citeup{19}$^{\hbox{\footnotesize ,}}$\citeup{50} Here, we
review some common diffusion-limited reactions with emphasis
on the effects of biased diffusion in two-species reactions.
Single-species reactions will also be briefly reviewed. 

The asymptotic diffusion-limited reaction behavior is associated
with diffusion being the slowest, `rate-determining' process in
the system.
For dimensions lower than the upper critical
dimension, a mean-field theory using the classical,
Smo\-lu\-chow\-ski-type
rate equations---see Kang and Redner\citeup{26}---breaks down since it
does not take into account effects of spatial inhomogeneities of
particle concentration. Specifically, the classical mean-field
theory predicts, in one dimension ($1d$), for the 
kinetics of single-species reactions---$\,A+A\to A$ and
$A+A\to \emptyset\;\, \hbox{(inert)}$---an asymptotic decay of the
concentration proportional to
$t^{-1}$ instead of the exact result
$t^{-1/2}$.\citeup{54}$^{\hbox{\footnotesize
-}}$\citeup{19}$^{\hbox{\footnotesize
,}}$\citeup{2}$^{\hbox{\footnotesize -}}$\citeup{67}
That diffusion is the mechanism leading to anomalously slow kinetics
can be illustrated for the annihilation reaction,
$A+A\to \emptyset  $, on a one-dimensional ($1d$)
lattice, as follows.  After a time $t$, a particle
will diffuse over a distance of order $\ell=\sqrt{Dt}$, where $D$
is the particle diffusion constant.
Let us assume that the
concentration at time $t$ is $C(t)$, so that the mean distance
between particles is $n(t)=1/C(t)$. In the limit $n\gg \ell$
it is reasonable to
assume that particles diffuse freely, i.e., perform
independent random walks, and there is no annihilation.
In the opposite limit, $n \ll \ell$, it is very unlikely for
a particle to actually diffuse over a distance
$\ell$ without being annihilated. Thus the variation of the particle
density with time will be determined by the `balance' of diffusion
and annihilation achieved when $n(t)\approx \ell$, and we get
$C(t)\sim t^{-1/2}$.

Once two particles annihilate, there develops a local fluctuation in
the density and correlation functions. The spatial extent of the
disturbance is $n(t)$ and therefore the time scale required to erase
this local fluctuation by diffusion is $n^2/D$. However, if
$n \approx \ell$, the next reaction event locally will occur on
the same time scale! Therefore, diffusion is ineffective in `mixing'
the particles and as a result the $1d$ dynamics is truly
fluctuation-dominated, non-mean-field. 
Similar heuristic arguments have been advanced\citeup{79} for the
$d$-dimensional case. 

Recent research has been focused on dimensions lower than the critical
dimension, where the mean-field description breaks down. In
particular, $1d$ problems are usually more amenable
to mathematical analysis; in several instances exact solutions
have been reported.\citeup{54}$^{\hbox{\footnotesize
,}}$\citeup{55}$^{\hbox{\footnotesize
,}}$\citeup{19}$^{\hbox{\footnotesize
,}}$\citeup{2}$^{\hbox{\footnotesize
-}}$\citeup{72}$^{\hbox{\footnotesize
,}}$\citeup{10}$^{\hbox{\footnotesize ,}}$\citeup{11}
Apart {}from their intrinsic interest, exact
results, when available, serve to test theoretical and
computational results and methods which are also applied to
more realistic problems.

Low-dimensional reaction-diffusion models and related simple
dynamical systems actually describe experimental phenomena as
diverse as electron-hole
recombination in semiconductors, soliton-antisoliton
dynamics in quasi-$1d$ systems, evolution of competing species in
biology, aerosol
dynamics, star formation, and
poly\-me\-ri\-za\-tion,\citeup{6}$^{\hbox{\footnotesize
,}}$\citeup{61}$^{\hbox{\footnotesize
,}}$\citeup{36}$^{\hbox{\footnotesize -}}$\citeup{31} etc.
Despite their inherent complexity and diversity such
systems share common features---some reviewed below. These
`universal' properties can be studied in simple stochastic
lattice-dynamics models.

Experimental investigations have been performed on quasi-$1d$
systems.\citeup{36}$^{\hbox{\footnotesize
,}}$\citeup{37}$^{\hbox{\footnotesize
,}}$\citeup{27}$^{\hbox{\footnotesize ,}}$\citeup{28}
In such systems
the hopping rate in the favorable direction is about $10^4$
times higher than in the plane perpendicular to that direction.
Therefore, the system can be regarded  as $1d$ to a good
approximation. Notably, hexagonal crystals of
TMMC\citeup{36}$^{\hbox{\footnotesize -}}$\citeup{65}
represent the best quasi-$1d$ system available experimentally. Excitons
propagate preferably along the $\hbox{Mn}^{2+}$ chains, as
compared to motion 
across chains; the latter distance is nearly three times
larger than the in-chain separation. The resulting hopping rate
between adjacent chains is eight orders of magnitude smaller
than the in-chain hopping rate. Typical values for the hopping
rates and lifetime are $10^{11}$-$10^{12}\,$s$^{-1}$ and $740\,\mu$s,
respectively. This large lifetime excludes natural decay of
the quasiparticles, which  would yield an 
exponential relaxation, so that the dynamics is dominated by
the quasiparticle 
reactions. It is also a good approximation to consider
the reaction time as instantaneous. Indeed, a pair of excitons
reacts in about $100\,$fs, i.e., at least one order of magnitude
smaller than the typical hopping time. Theoretical studies of
the effects of finite probability of reactions have been reported
in the literature.\citeup{12}$^{\hbox{\footnotesize
,}}$\citeup{56}$^{\hbox{\footnotesize -}}$\citeup{DbADZ}
Next, we focus on analytical results for single- and two-species
reactions.

\textheight=7.8truein
\setcounter{footnote}{0}
\renewcommand{\thefootnote}{\alph{footnote}}

\section{Some common diffusion-limited reactions}
\noindent
The kinetics of diffusing-particle single-species $1d$ systems
reacting via the
processes of coalescence ($A+A\to A$) or annihilation ($A+A\to
\emptyset $) on particle encounters, is now well understood. For
these two basic processes
the concentration decays for large times according to
$C(t)\sim t^{-1/2}$ (in $1d$). In general,
$C(t)\sim t^{-d/2}$ for $d < 2$, with a logarithmic correction
factor at $d=d_c=2$,
and $C(t)\sim t^{-1}$ for $d>2$.\citeup{79}$^{\hbox{\footnotesize
,}}$\citeup{54}$^{\hbox{\footnotesize
,}}$\citeup{2}$^{\hbox{\footnotesize
-}}$\citeup{9}$^{\hbox{\footnotesize
,}}$\citeup{47}$^{\hbox{\footnotesize
 -}}$\citeup{5}$^{\hbox{\footnotesize
,}}$\citeup{17}$^{\hbox{\footnotesize
,}}$\citeup{44}$^{\hbox{\footnotesize
,}}$\citeup{45}
Here $d_c$ denotes the upper
critical dimension. The steady state, assuming
additional processes, e.g., particle creation or back-reaction,
has also been
studied by several authors.\citeup{6}$^{\hbox{\footnotesize
,}}$\citeup{5}$^{\hbox{\footnotesize
,}}$\citeup{14}$^{\hbox{\footnotesize
,}}$\citeup{16a}$^{\hbox{\footnotesize
,}}$\citeup{45}
In particular,
R\'acz\citeup{61} obtained exact results by mapping interfaces
of the kinetic Glauber-Ising model to particles in the
$A+A \leftrightarrow \emptyset$ system. Amar and Family\citeup{81}
pointed out that the Ising-model to reaction-dynamics mapping
should be used with caution owing to the presence of subtle
correlations in the initial conditions for non-zero values of
the magnetization. Multiparticle $kA\to \emptyset $ reactions,
and other variants of single-species
reactions have also attracted
attention.\citeup{59}$^{\hbox{\footnotesize
,}}$\citeup{40}$^{\hbox{\footnotesize
,}}$\citeup{23}$^{\hbox{\footnotesize
,}}$\citeup{58}$^{\hbox{\footnotesize ,}}$\citeup{24}
For $k>3$, a mean-field approximation applies in
$1d$. For the borderline case of $k=3$,
Monte-Carlo simulations\citeup{82} and refined mean-field
treatment\citeup{83} have recently succeeded in demonstrating the
logarithmic correction in the concentration decay.

The kinetic behavior of two-species reactions is far
richer than that of the
single-species reactions. Three two-species models belonging
to different universality classes are discussed
below. Although there exist other variants of two-species
models,\citeup{38}$^{\hbox{\footnotesize
,}}$\citeup{24}$^{\hbox{\footnotesize
,}}$\citeup{4}$^{\hbox{\footnotesize -}}$\citeup{32}
we restrict the present survey to these few examples.
In the `standard' $AB$ model particles hop isotropically 
on a $d$-dimensional lattice to any of their
neighboring sites. On encounters, $AB$ particle pairs annihilate
while for $AA$ and $BB$ pairs the two particles `bounce off'
each other. This
models a hard-core interaction whereby each site can
be occupied by at most one particle.

The `standard' $AB$ model shows the non-mean-field concentration decay
$C(t)\sim t^{-d/4}$ for $d<4$ and the mean-field decay 
$C(t)\sim t^{-1}$ for $d > d_c=4$, for equal concentrations of
the two species.\citeup{79}$^{\hbox{\footnotesize
,}}$\citeup{44}$^{\hbox{\footnotesize
,}}$\citeup{10}$^{\hbox{\footnotesize
,}}$\citeup{11}$^{\hbox{\footnotesize
,}}$\citeup{24}$^{\hbox{\footnotesize
,}}$\citeup{25}$^{\hbox{\footnotesize -}}$\citeup{70}
For unequal (initial) concentrations, the
concentration of the
minority species decays exponentially as $C(t)\sim
\exp{(-\lambda\sqrt{t})}$ for $d=1$, $C(t)\sim\exp{(-\lambda t/\ln
t)}$ for $d=2$, and $C(t)\sim\exp{(-\lambda t)}$ for $d\ge
3$.\citeup{10}$^{\hbox{\footnotesize
,}}$\citeup{11}$^{\hbox{\footnotesize
,}}$\citeup{25} The 
seminal works by Bramson and Lebowitz summarize the
exactly known results for the $d\ge 1$ `standard' $AB$
model.\citeup{10}$^{\hbox{\footnotesize
,}}$\citeup{11}

A `driven' $1d$ version of the $AB$ model was recently
introduced by Janowsky\citeup{21}: particles are constrained to
hop in one direction only.
When combined with hard-core interactions, the biased-hopping dynamics
yields a new universality class.
Phenomenological considerations\citeup{21}$^{\hbox{\footnotesize
,}}$\citeup{22}  relate this reaction model
to the asymmetric simple
exclusion process, for which the noisy Burgers' equation
represents the continuum limit.
Numerical simulations and the above connection to the noisy
Burgers' equation suggest the concentration-decay exponent $1\over
3$, compared to $1\over 4$ of the
`standard' $AB$ model, in $1d$. Regarding
the spatial domain structure, there is only a qualitative
agreement in the literature,\citeup{22}$^{\hbox{\footnotesize
,}}$\citeup{20} including
phenomenological considerations and numerical simulations. 

\section{Exactly solvable `sticky' two-species annihilation model}
\noindent
A third variant of the $AB$ model in $1d$ has been introduced recently by
Kra\-piv\-sky.\citeup{30}$^{\hbox{\footnotesize
,}}$\citeup{31}$^{\hbox{\footnotesize ,}}$\citeup{29}
In this system same-species particles interact by sticking together.
Same-species particle clusters coalesce on encounters---no hard-core
interaction is
present---and they form a larger cluster with size
given by the total number of particles in the merging
clusters. When two clusters of different-species particles meet
$AB$-annihilation takes place. The 
resulting cluster has particle number given by the particle-number
difference of the reacting clusters, while 
the species is that of the `parent' cluster with the
larger number of particles. Numerical simulations and
scaling\citeup{71} suggest $d_c=2$.

Synchronous-updating, cellular-automaton-like,  `sticky-particle'
lattice dynamics models
have been considered recently. Exact results have been obtained for
`sticky' diffusion-limited
single-species,\citeup{52} and single-
and two-species\citeup{54}$^{\hbox{\footnotesize ,}}$\citeup{55}
reactions. They have been treated by a
novel approach which yields a unified exact solution of
the coagulation and annihilation single-species reactions
and the `sticky' two-species reaction defined above.\citeup{54}
The approach combines and extends several techniques 
developed in charge-coagulation model 
studies\citeup{74}$^{\hbox{\footnotesize -}}$\citeup{48}
and earlier single-species reaction 
studies\citeup{50}$^{\hbox{\footnotesize ,}}$\citeup{52}
of cellular-automaton-type models. A combinatorial
argument is then utilized to relate the `cluster history' at time $t$
to the set of compatible initial configurations (at
$t=0$).\citeup{31} Exact solvability for the two-species reaction
is based on the linearity of  certain dynamical evolution 
equations, supplemented by a mathematical development yielding
a closed form for a non-trivial double-sum.\citeup{54}

For the single-species reactions the expected power-law
decay with exponent $1\over 2$ was confirmed, regardless of
the hopping rate bias. Moreover, for large times (only) the effect
of the hopping-rate anisotropy can be fully absorbed in the
time dependence, i.e., it
only changes the value of the diffusion constant. The method
also allows one to extend the exact relation,
$C_A(t;2p)=2C_\emptyset(t;p)$, where the subscripts represent
the reaction outcome and $p$ is the initial particle concentration,
to the biased-hopping case. Similar `duality' relations were 
found in other formulations of  $A+A \to A$ or $\emptyset$
models,\citeup{2}$^{\hbox{\footnotesize ,}}$\citeup{72}
with microscopically different dynamical rules, although for
the general case they only apply asymptotically for large times.

For the `sticky' two-species reaction, we were able to derive
several exact results. The concentration decay, for
equal initial concentrations of both species, is $\sim t^{-1/4}$ as
in the `standard' $AB$ case. Furthermore, the exponent value 
$1\over 4$  is not
affected here by the hopping-rate bias as it is for the `driven'
hard-core $AB$
model. For unequal initial concentration of the two species, the
asymptotic approach to the constant concentration difference
is $t^{-3/2}$. This clearly contrasts with the exponential
dependence of the `standard' $AB$ model. Thus, for the 
multiple-occupancy, sticky-particle `bosonic' model the effects of  the
hopping bias are not fundamentally important. In fact, in the large-time
asymptotic regime the changes due to bias can be
fully absorbed in the diffusion constant value. This behavior is
quite different {}from that of the single-occupancy, hard-core
`fermionic' models.

To summarize, we have reviewed some common diffusion-limited
reactions with 
emphasis on exact results and our recent exact solution of the
`sticky' two-species model. Many open problems challenge
researchers in this field.
The few exact results available yield  insight on the
physical problem and provide a guide for approximate analytical
and numerical
work. We have emphasized the differences between the three two-species
reactions considered which actually belong to different
universality classes.

\nonumsection{References}

\end{document}